\documentclass[aps,prl,reprint,superscriptaddress]{revtex4-1}

\usepackage{graphicx}
\usepackage{amsmath}
\usepackage{textcomp} 
\usepackage{xcolor}
\usepackage{ulem}

\begin{document}

\title{Interface-driven spin-torque ferromagnetic resonance by Rashba coupling at the interface between non-magnetic materials}

\author{M.~B.~Jungfleisch}
\email{jungfleisch@anl.gov}
\affiliation{Materials Science Division, Argonne National Laboratory, Argonne IL 60439, USA}

\author{W.~Zhang}
\affiliation{Materials Science Division, Argonne National Laboratory, Argonne IL 60439, USA}

\author{J.~Sklenar}
\affiliation{Materials Science Division, Argonne National Laboratory, Argonne IL 60439, USA}
\affiliation{Department of Physics and Astronomy, Northwestern University, Evanston IL 60208, USA}

\author{W.~Jiang}
\affiliation{Materials Science Division, Argonne National Laboratory, Argonne IL 60439, USA}

\author{J.~E.~Pearson}
\affiliation{Materials Science Division, Argonne National Laboratory, Argonne IL 60439, USA}

 \author{J.~B.~Ketterson}
\affiliation{Department of Physics and Astronomy, Northwestern University, Evanston IL 60208, USA}

\author{A.~Hoffmann}
\affiliation{Materials Science Division, Argonne National Laboratory, Argonne IL 60439, USA}

\date{\today}

\begin{abstract}

The Rashba-Edelstein effect stems from the interaction between the electron's spin and its momentum induced by spin-orbit interaction at an interface or a surface. It was shown that the inverse Rashba-Edelstein effect can be used to convert a spin- into a charge current. Here, we demonstrate that a Bi/Ag Rashba interface can even drive an adjacent ferromagnet to resonance. 
We employ a spin-torque ferromagnetic resonance excitation/detection scheme which was developed originally for a bulk spin-orbital effect, the spin Hall effect. In our experiment, the direct Rashba-Edelstein effect generates an oscillating spin current from an alternating charge current driving the magnetization precession in a neighboring permalloy (Py, Ni$_{80}$Fe$_{20}$) layer. Electrical detection of the magnetization dynamics is achieved by a rectification mechanism of the time dependent multilayer resistance arising from the anisotropic magnetoresistance.


\end{abstract}
\maketitle

Conventional spintronics relies on the exchange interaction between conduction electrons on one side and localized spins in magnetic materials on the other side \cite{Zutic}. Stimulated by the experimental demonstration of spin- to charge current conversion using bulk spin Hall effects (SHE), these kind of spin-orbital phenomena were actively investigated in the last decade and opened up the door to the research field of spin-orbitronics \cite{Dyakonov,Hirsch,Hoffmann,Kajiwara}. SHEs can be investigated by means of spin-current injection from a ferromagnet (FM) into materials with large spin-orbit coupling, usually normal metals (NM) such as Pt or Pd \cite{Wei_PRB}, and sensing the generated voltage generated by means of the inverse spin Hall effect (ISHE) \cite{Mosendz,Mosendz_PRB,Wei_PRL,Saitoh,Azevedo,Wei_JAP,Jungfleisch,Jungfleisch_PRB}. Other interesting applications of SHEs are the effective magnetization switching of nanomagnets or the movement of domain walls \cite{STT,Mesoscale_mag,Wanjun_Science}. Furthermore, the ferromagnetic linewidth modulation as well as the excitation of spin waves and ferromagnetic resonance by SHE was demonstrated in ferromagnetic metals and insulators \cite{Liu_ST-FMR,Joe_PRB,Klein_PRL,Wu,Mellnik_ST-FMR}. The SHE is a bulk effect occurring within a certain volume of the NM determined by the spin-diffusion length. The conversion efficiency can be expressed by a material-specific parameter, the spin Hall angle $\gamma_\mathrm{SHE}$ \cite{Hoffmann}.

\begin{figure}[t]
\includegraphics[width=1\columnwidth]{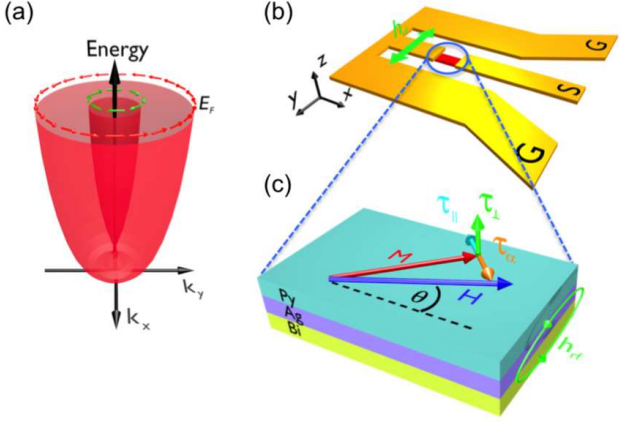}
\caption{\label{Fig1} (Color online) (a) Dispersion curves of a 2D electron gas are spin-split due to the REE. (b) Scheme of the ST-FMR experimental setup. (c) ST-FMR mechanism in Py/Ag/Bi multilayers. The alternating RF current drives an Oersted field $h_\mathrm{RF}$ exerting a field-like torque $\tau_\mathrm{\perp}$ on the magnetization $M$. At the same time a oscillatory transverse spin accumulation at the Py/Ag interface generated at the Ag/Bi interface by the REE exerts a damping-like torque $\tau_\mathrm{\vert\vert}$ on the magnetization.} 
\end{figure}

Very recently, it has been shown that the inverse Rashba-Edelstein effect (IREE) can also be used for transformation of a spin- into a charge current \cite{Sanchez_Nat_com,Wei_IREE,Sangiao,Ando_IREE}. The IREE is the inverse process to the Rashba-Edelstein effect (REE) \cite{Edelstein}. The REE originates from spin-orbit interaction in a 2D electron gas at interfaces or surfaces, which effectively produce a steady non-equilibrium spin polarization from a charge current driven by an electric field. The Hamiltonian of this interaction is given by \cite{Sanchez_Nat_com}: ${H_\mathrm{R}=\alpha_\mathrm{R}(k\times \hat{e}_\mathrm{z}) \cdot\sigma,}$
%
where $\alpha_\mathrm{R}$ is the Rashba coefficient, $\hat{e}_\mathrm{z}$ is the unit vector in $z$-direction [see Fig.~\ref{Fig1}(b,c)] and $\sigma$ is the vector of Pauli matrices. As a result of this interaction the dispersion curves of the 2D electron gas are spin-split if $\alpha_\mathrm{R} \neq 0$, as illustrated in Fig.~\ref{Fig1}(a). Analogous to the spin Hall angle, the spin- to charge current interconversion parameter can be defined as \cite{Sanchez_Nat_com}:
\begin{equation}
\label{Rashba_parameter}
{\lambda_\mathrm{REE}=\alpha_\mathrm{R}\tau_\mathrm{S}/\hbar,}
\end{equation}
where $\tau_\mathrm{S}$ is the effective relaxation time describing the ratio between spin injection and spin-momentum scattering and $\hbar$ is the reduced Planck constant.
The spin-split 2D electron gas dispersions and Fermi contours of many Rashba surfaces and interfaces have been investigated spectroscopically \cite{Ast}. In general, large Rashba couplings occur at interfaces between heavy elements with strong spin-orbit interaction (such as Bi, Pb, and Sb) and other non-magnetic materials with small spin-orbit coupling such as Ag, Au, and Cu \cite{Ast, Koroteev}. Even though, the interaction between a charge current and a non-zero spin density at a Rashba interface has been demonstrated by injection of a spin-pumping driven spin current at ferromagnetic resonance, the reverse process remains to be explored experimentally until now.

Here, we demonstrate that a Bi/Ag Rashba interface can drive spin-torque ferromagnetic resonance (ST-FMR) in an adjacent ferromagnetic layer. We interpret our results in terms of an excitation by the direct REE, which drives an oscillating spin current from an alternating charge current that scatters of the Rashba interface (Ag/Bi). The generated spin current excites the magnetization precession in a neighboring permalloy (Py, Ni$_{80}$Fe$_{20}$) layer by the spin-transfer torque effect \cite{Liu_ST-FMR,Slonczewski}. The precessional magnetization leads to resistance oscillations on account of the anisotropic magnetoresistance (AMR) of Py. The mixing between the applied alternating current and the oscillating resistance allows for a direct voltage detection of the induced magnetization dynamics \cite{Liu_ST-FMR,Mellnik_ST-FMR}. Injecting an additional DC current to the sample results in an additional spin current generation due to the REE which enables to manipulate the ferromagnetic resonance linewidth by exerting a torque on the magnetization.


We fabricated the devices using magnetron sputtering and photolithography. The multilayers were prepared in the shape of $30\times5$~$\mu$m$^2$ stripes using lithography and lift-off on intrinsic Si substrates with $300$-nm thick thermally grown SiO$_2$. Four different types of multilayers were deposited using magnetron sputtering: Py, Py/Bi, Py/Ag and Py/Ag/Bi. In the case of the Py/Ag/Bi systems, the Ag thickness was $t_\mathrm{Ag}$ = 2, 4, 6, 10, 15 nm, the Py thickness $t_\mathrm{Py}= 9$~nm and the Bi thickness $t_\mathrm{Bi}= 4$~nm. The control samples feature a Py thickness of 7 nm, Ag thickness 6 nm and Bi thickness 4 nm. In a subsequent process step, the coplanar waveguide (CPW) was fabricated on top of the multilayers. Figure~\ref{Fig1}(b) illustrates the experimental setup. A bias-T is used to apply a microwave signal and to detect the rectified DC voltage at the same time. The applied microwave power is kept constant at $+10$~dBm, unless otherwise mentioned. An in-plane magnetic field is applied at an angle of $\theta = 45^\circ$ [see illustration in Fig.~\ref{Fig1}(b,c)]. While sweeping the external magnetic field the DC voltage is detected by a lock-in amplifier with an amplitude modulation at 3~kHz. All measurements were performed at room temperature.

\begin{figure}[b]
\includegraphics[width=1\columnwidth]{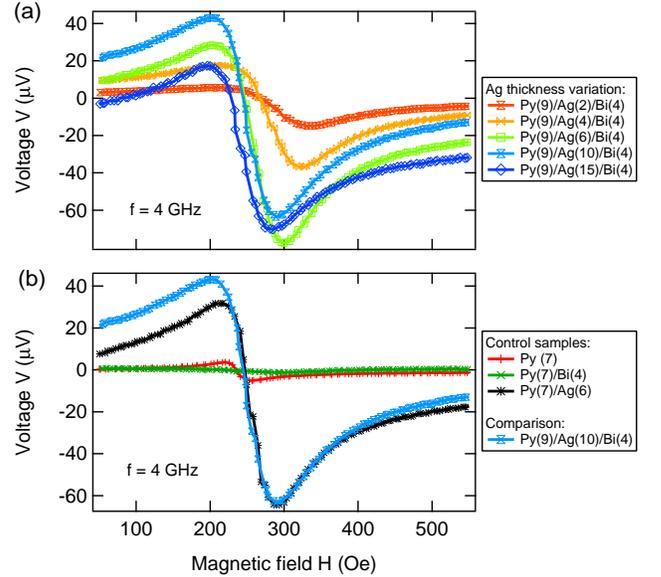}
\caption{\label{Fig2} (Color online) Spectra of REE-driven ST-FMR measured at a frequency of 4 GHz and an applied microwave power of +10 dBm. 
Thickness in brackets given in nm. (a)  Ag thickness dependence of the resonance signal. (b) Comparison between control samples and Py(9)/Ag(10)/Bi(4).} 
\end{figure}

Figure~\ref{Fig2} shows typical spectra at an excitation frequency of $f=4$~GHz. Let's first discuss the trilayers [Fig.~\ref{Fig2}(a)]. In our experiment, magnetization dynamics is excited simultaneously by the Oersted field as well as by the REE which generates an oscillating spin current from the alternating charge current driving the magnetization precession in the neighboring permalloy layer when the condition of ferromagnetic resonance is fulfilled, 
\begin{equation}
\label{Kittel}
{f= \frac{\vert\gamma\vert}{2\pi} \sqrt{H_\mathrm{}(H_\mathrm{}+4\pi M_\mathrm{eff})}.}
\end{equation}
Here, $M_\mathrm{eff}$ is the effective magnetization and $\vert\gamma\vert$ is the gyromagnetic ratio. Electrical detection of the magnetization dynamics is achieved by a rectification mechanism of the time dependent multilayer resistance arising from the AMR of Py. A rectification by spin pumping and IREE is a secondary effect in our experiment \cite{Liu_ST-FMR}. As apparent from Fig.~\ref{Fig2}(a), the Py/Ag/Bi samples exhibit a superimposed symmetric and antisymmetric Lorentzian lineshape. The smallest Ag interlayer thickness of 2 nm shows the largest symmetric contribution, but the smallest absolute signal. With increasing $t_\mathrm{Ag}$ the signal tends to be more antisymmetric and the absolute value increases. The control samples are depicted in Fig.~\ref{Fig2}(b). The pure Py sample features a small, antisymmetric Lorentzian signal due to a rectification by AMR. The Py/Bi sample exhibits a very small, mostly symmetric signal. Py/Ag features a reasonably large antisymmetric signal: The Ag layer is beneficial for the absolute voltage because a larger Oersted field is generated in the Py layer resulting in a higher AMR signal manifested in a substantial antisymmetric lineshape.

\begin{figure}[t]
\includegraphics[width=1\columnwidth]{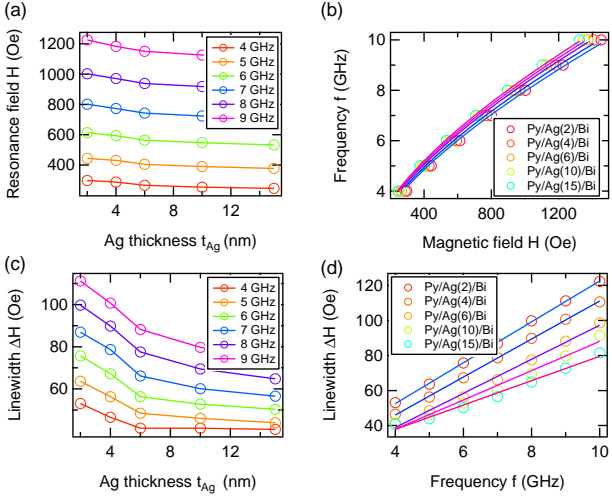}
\caption{\label{Fig3} (Color online) (a) Resonance at various excitation frequencies for different Ag thicknesses. (b) Dispersion measured for different Ag interlayer thicknesses, $t_\mathrm{Py}=9$~nm, $t_\mathrm{Bi}=4$~nm. A fit to Eq.~(\ref{Kittel}) confirms the excitation of ferromagnetic resonance; shown as solid lines. (c) Evolution of the FMR linewidth with $t_\mathrm{Ag}$ at different excitation frequencies. (d) Determination of Gilbert damping parameter $\alpha$. Solid lines show a fit  to Eq.~(\ref{damping}).} 
\end{figure}

Figures~\ref{Fig3}(a) and (c) illustrate how the resonance field and linewidth alter for different Ag interlayer thicknesses at various excitation frequencies. 
The excitation of ferromagnetic resonance is confirmed by a fit to Eq.~(\ref{Kittel}), see Fig.~\ref{Fig3}(b). Furthermore, the data shown in Fig.~\ref{Fig3}(d) is governed by a linear dependence between linewidth $\Delta H$ and the excitation frequency $f$: 
\begin{equation}
\label{damping}
\Delta H(f) = \Delta H_0 + 4\pi f \frac{\alpha}{\vert\gamma\vert},
\end{equation}
where $\Delta H_0$ is the inhomogeneous linewidth broadening given by the zero-frequency intercept and $\alpha$ is the Gilbert damping parameter. This confirms the excitation of FMR in our samples.

The magnetization dynamics in a macrospin model is governed by a modified Landau-Lifshitz-Gilbert equation \cite{Mellnik_ST-FMR}:
\begin{equation}
\label{ST-FMR}
\begin{split}
\frac{\mathrm{d}\hat{m}}{\mathrm{d}t}=-\vert\gamma\vert \hat{m}\times\vec{H}_\mathrm{eff} + &  \alpha  \hat{m}\times\frac{\mathrm{d}\hat{m}}{\mathrm{d}t}  +  \vert\gamma\vert \tau_{\parallel} \hat{m}\times(\hat{y} \times \hat{m})\\+ \vert\gamma\vert \tau_\perp \hat{y}\times \hat{m},
\end{split}
\end{equation}
where $\hat{m}$ is the magnetization direction, $H_\mathrm{eff}$ is the effective magnetic field, $\tau_{||}$ and $\tau_{\perp}$ are the two acting torque components, and the coordinate system ($\hat{x}, \hat{y}, \hat{z}$) is defined as shown in Fig.~\ref{Fig1}(b,c).

\begin{figure}[b]
\includegraphics[width=1\columnwidth]{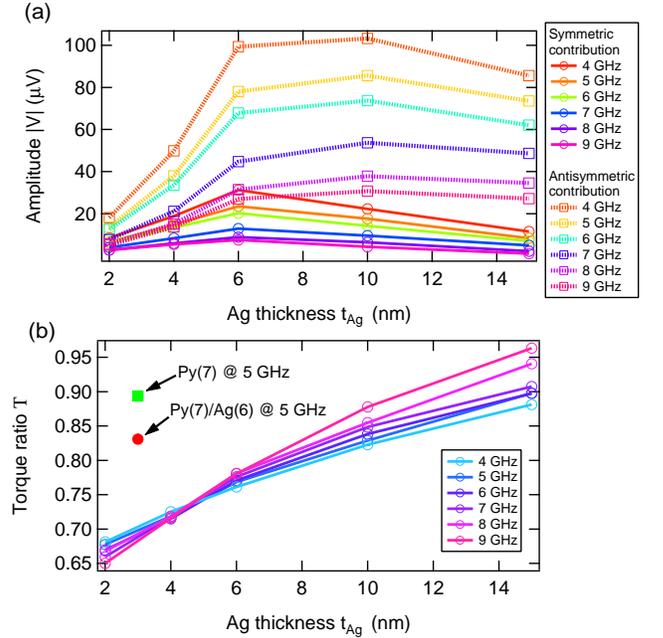}
\caption{\label{Fig4} (Color online) (a) Deconvoluted symmetric and antisymmetric contribution to DC voltage amplitude for various Ag interlayer thicknesses. (b) Ratio $\mathrm{T}= V_\mathrm{antisymm}/(V_\mathrm{antisymm}+V_\mathrm{symm})$ as function of $t_\mathrm{Ag}$ for various frequencies.} 
\end{figure}

The two vector components of the current-induced torque $\tau_{||}, \tau_{\perp}$ can be related to the amplitudes of the symmetric and antisymmetric components of the resonance lineshape \cite{Mellnik_ST-FMR}: (1) An in-plane component $\tau_\mathrm{\vert\vert} \sim \hat{m}\times(\hat{y}\times\hat{m})$ results in a symmetric contribution and (2) an out-of-plane component $\tau_\mathrm{\perp} \sim \hat{y}\times\hat{m}$ results in an antisymmetric contribution, see Fig.~\ref{Fig1}(c) \cite{Mellnik_ST-FMR}. 
Figure~\ref{Fig4}(a) illustrates the Ag thickness dependence of the amplitudes of both contributions, respectively, as a function of the driving RF frequency. We observe the following trend: The amplitudes increase with increasing $t_\mathrm{Ag}$ up to $t_\mathrm{Ag}\approx 7$~nm. At larger thicknesses, the antisymmetric contribution (dashed lines) remains constant up to $t_\mathrm{Ag}\approx 10$~nm before it decreases. The symmetric contribution, however, peaks at $t_\mathrm{Ag}\approx 7$~nm and reduces afterwards. In order to highlight this observation we plot a torque-ratio equivalent $\mathrm{T} = V_\mathrm{antisymm}/(V_\mathrm{antisymm}+V_\mathrm{symm})$ as a function of $t_\mathrm{Ag}$ in Fig.~\ref{Fig4}(b). Clearly, the symmetric contribution to the lineshape is greatest at a lower Ag thickness and becomes negligible for larger $t_\mathrm{Ag}$.  The reason for this trend is the larger Oersted field produced in samples with a thicker Ag layer and, thus, a larger out-of-plane torque contribution $\tau_\mathrm{\perp}$. As is apparent from Fig.~\ref{Fig4}(b), this trend is independent on the excitation frequency. We also show the ratio of the control samples Py/Ag and Py in the same plot as a red dot and a green square, respectively. 

We interpret our observations in the following way: If the observed increase of the symmetric component ($\sim \tau_\mathrm{\vert\vert}$) with respect to the antisymmetric component ($\sim \tau_\mathrm{\perp}$) was caused by the SHE in Ag, we should observe the same ratio for the control sample Py/Ag. As is apparent from Fig.~\ref{Fig4}(b), this is not the case. Since Ag features a long spin-diffusion length of $\sim 300$~nm \cite{Ag-diffusion}, it would also be possible that the SHE in Bi generates a spin current which diffuses through the Ag layer. However, since the generated voltage for the control sample Py/Bi is negligibly small, see Fig.~\ref{Fig2}(b), this mechanism can also be ruled out. We conclude that the magnetization dynamics in our Py/Ag/Bi samples is driven by an interfacial charge-spin conversion due to the REE. 

\begin{figure}[t]
\includegraphics[width=1\columnwidth]{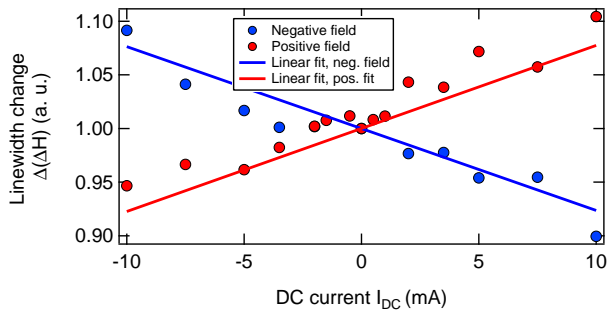}
\caption{\label{Fig5} (Color online) Manipulation of the FMR linewidth by a simultaneous injection of an electrical DC current. Py(15)/Ag(4)/Bi(4), $f = 4$~GHz, $P_\mathrm{RF} = +2$~dBm.} 
\end{figure}

According to the spin-torque theory \cite{Slonczewski}, an additional spin current injected into the FM layer will increase or decrease the effective magnetic damping, i.e., the linewidth, depending on its relative orientation with respect to the magnetization \cite{Liu_ST-FMR,Wu}. Since Ag features a very small spin Hall angle \cite{Hailong} and our Bi layer is almost non-conducting \cite{Wei_IREE}, the demonstration of the ferromagnetic linewidth manipulation by an additional DC current injection would be an independent manifestation of charge- to spin current conversion by the REE.
Figure~\ref{Fig5} shows unambiguously that it is indeed possible to manipulate the resonance lineshape if an additional DC current is injected into the sample. For this purpose a rather small RF power of +2 dBm is chosen. Apparently, for a positive magnetic field polarity, a positive DC current leads to an enhanced linewidth, i.e., a damping enhancement. In contrast, a negative current leads to a decreased linewidth, i.e., a damping reduction. Reversing the field polarity results in an opposite trend. We find a relative linewidth change of $0.8\%$~mA$^{-1}$.

Although it isn't physical to speak of a thickness in case of an interface effect, it is still possible to adapt a lineshape analysis approach which was presented originally in Ref.~\cite{Liu_ST-FMR} to relate the spin Hall angle to the ratio symmetric/antisymmetric components of the resonance lineshape. We can estimate a spin Hall angle equivalent $\gamma^{*}$ if we hypothetically assume that the charge-spin conversion process was a bulk- rather than an interface-driven effect \cite{Liu_ST-FMR}: 
\begin{equation}
\label{SH_angle}
\gamma^{*} = \frac{S}{A}\frac{e \mu_0 M_\mathrm{S}t_\mathrm{Py} t_\mathrm{NM} }{\hbar} \sqrt{1+\frac{4\pi M_\mathrm{eff}}{H_\mathrm{}}}.
\end{equation}
Here, $t_\mathrm{NM}$ is the non-magnetic layer thickness. We find the spin Hall angle equivalent to be $\gamma^{*}\approx18$\% for our Py/Ag/Bi samples, exceeding most paramagnetic metals. In our previous work we determined the REE conversion parameter $\lambda_\mathrm{REE}\approx 0.1$~nm \cite{Wei_IREE}. Using the relation $\lambda_\mathrm{REE}= 1/2 d \gamma^{*}$, where $d$ is the \textit{interface layer} thickness \cite{Sanchez_Nat_com}, we obtain $d\approx 1$~nm, which is a reasonable estimate.

In summary, we demonstrated the conversion of a charge- into a spin current by Rashba coupling of interface states by adapting a spin-torque ferromagnetic resonance excitation/detection technique. The Ag thickness dependence clearly demonstrates that the spin dynamics in the adjacent Py layer is driven by an interface-generated spin-polarized electron current that exerts a torque on the magnetization rather than a bulk effect such as the spin Hall effect. Our conclusions are further validated by a FMR linewidth modulation due to the spin current injection by applying an additional DC charge current to the sample stack. Our results will stimulate experimental and theoretical endeavors to explore novel interface- and surface-driven spin-orbital phenomena for the efficient excitation of magnetization dynamics.

We thank Roland~Winkler for illuminating discussions. This work was supported by the U.S. Department of Energy, Office of Science, Materials Science and Engineering Division. Lithography was carried out at the Center for Nanoscale Materials, an Office of Science user facility, which is supported by DOE, Office of Science, Basic Energy Science under Contract No. DE-AC02-06CH11357.

\end{document}